\documentclass[12pt]{article}
\usepackage{pazha}
\usepackage{graphicx}
\usepackage{lscape}
\usepackage{amssymb}
\usepackage{amsmath}
\usepackage{graphicx}
\usepackage[T2A]{fontenc}
\graphicspath{{E:/!Gravitational_Lensing/!Images/astro_ph/}}
\tightenlines
\def\ÐÍ{$\pm$}

\begin{document}

{\it Will be published in ``Astronomy Letters'', 2011, v.37, N4, pp. 233-247}

\bigskip

\title{\bf Modeling the Images of Relativistic Jets Lensed by Galaxies with Different Mass Surface Density Distributions}

\author{\bf T.I.Larchenkova\affilmark{1,*}, A.A.Lutovinov\affilmark{2,**}, and N.S.Lyskova\affilmark{3}}

\affil{ {\it Astro Space Center, Lebedev Physical Institute, Russian
Academy of Sciences, Profsoyuznaya str., 84/32, Moscow, 117997
Russia}$^1$\\
{\it Space Research Institute, Russian Academy of
Sciences, Profsoyuznaya str., 84/32, Moscow, 117997 Russia}$^2$\\
{\it Moscow Institute of Physics and Technology, Institutskii per. 9, Dolgoprudnyi, Moscow obl., 141700 Russia}$^3$}
\vspace{2mm}
\received{22 June 2010}

\sloppypar
\vspace{2mm}
\noindent

The images of relativistic jets from extragalactic sources produced by gravitational lensing by galaxies with different mass surface density distributions are modeled. In particular, the following models of the gravitational lens mass distribution are considered: a singular isothermal ellipsoid, an isothermal ellipsoid with a core, two- and three-component models with a galactic disk, halo, and bulge. The modeled images are compared both between themselves and with available observations. Different sets of parameters are shown to exist for the gravitationally lensed system B0218+357 in multicomponent models. These sets allow the observed geometry of the system and the intensity ratio of the compact core images to be obtained, but they lead to a significant variety in the Hubble constant determined from the modeling results.

\vspace{10mm}

\vfill

$^*$ e-mail: tanya@lukash.asc.rssi.ru

$^{**}$ e-mail: aal@hea.iki.rssi.ru

\clearpage

\section*{INTRODUCTION}

\vskip -5pt

The formation and physical properties of largescale
relativistic plasma jets from active galactic nuclei,
quasars, and radio galaxies are among the topical
questions of modern extragalactic astronomy. The
relativistic jet from the giant elliptical galaxy M87
whose distance is 16.7 Mpc has been studied most
extensively owing to its close spatial location. The
question of whether the observed one-sided jet is
the result of its motion directed toward the observer,
whereby the jet knots move with a speed close to the
speed of light and their emission is more intense than
that of the counterjet due to the Doppler effect, or the
emission is actually anisotropic remains without an
answer so far.

Using the gravitational lensing of distant galactic
nuclei, quasars, and compact regions of radio galaxies
with large-scale relativistic jets not only gives the
observers a unique opportunity to see such astrophysical
objects but also will, possibly, allow individual
features of their jets, for example, the counterjet
unobservable in the absence of lensing, to be studied
in future. It is well known that the gravitational
lensing of a compact source with a relativistic jet can give rise to multiple images of the source itself and its
extended jet. Such gravitationally lensed systems are
actually observed at present. MG 1131+0456 (Hewitt
et al. 1988), PKS 1830–211 (Nair et al. 1993),
and B0218+357 (Patnaik et al. 1993, 1995) are the
brightest of them in the radio band. Therefore, modeling
the images of these sources and investigating
their behavior with time becomes a topical task. The
planned launches of very long baseline space interferometers
with a high angular resolution make this
task important and timely. Comparison of the simulations
of gravitationally lensed sources with observational
data will provide additional information both
about the physical radiation processes and collimation
mechanisms and about the mass density distribution
in the lens object.

The source B0218+357 is of particular interest
among the listed gravitationally lensed systems
with relativistic jets. The presence of a largescale
jet with an extent of  $\sim$ 1 Mpc observed in
the radio band, its considerable distance from other
extragalactic sources, and the accurately measured
time delay between its images (Biggs et al. 1999;
Cohen et al. 2000) make the study of this object
attractive, among other things, for an independent
determination of the Hubble constant: first, the
source lies at a high redshift at which the peculiar velocities are lower than the velocities of the Hubble
law; second, measuring the Hubble constant from the
time delay is a direct method of measurement, i.e., the
geometric scale of the gravitationally lensed system is
measured directly (Narayan and Bartelmann 1996).
Note that the lens in this system is a spiral galaxy
(Browne et al. 1993). This is a rare event per se,
because most of the gravitational lenses known to
date are elliptical galaxies. Apart from Â0218+357,
only nine systems in which the lens is a a spiral galaxy
are known to date (Feron et al. 2009).

The shape distortion and the appearance of multiple
images of a relativistic jet in the case of its gravitational
lensing by a galaxy are of interest in the study.
In particular, how important is it to take into account
the multicomponent structure of the spiral galaxy
when it lenses the jet? How will the noncoaxiality of
the galactic components affect the picture of lensing?
Under which conditions for gravitational lensing of
the jet do ring structures emerge in the image? Is
the assumption that the observed ``radio rings'' are the
images of the lensed jet justified? Investigating these
questions is the goal of our paper.

It seems natural to investigate the gravitational
lensing of a jet separately by early- and late-type
galaxies. Therefore, we will consider models of the
mass surface density distribution in elliptical and spiral
lens galaxies. In the case of elliptical galaxies,
either a homeoidal elliptical mass surface density
distribution or an elliptical effective lensing potential
are considered (see, e.g., Kassiola and Kovner 1993;
Kormann et al. 1994). For a homeoidal density distribution,
all surfaces of equal density are represented by
concentric, similar, and identically oriented ellipsoids
(King 2002). In this paper, when modeling both elliptical
and spiral galaxies, we use models of the mass
surface density distribution rather than the effective
potential, because the models with an elliptical potential
are inapplicable at high ellipticities (Kormann
et al. 1994).

For spiral lens galaxies, Keeton and Kochanek
(1998) suggested using models including their multicomponent
structure, namely, their disk, bulge, and
halo. In this paper, when modeling the gravitational
lensing of a relativistic jet by a spiral galaxy, we
will use the models proposed in the paper mentioned
above. The choice of the models is dictated by the
fulfilment of the following main requirements: an adequate
physical description of the observed phenomena
and the possibility of representing the lens equation
in analytic form (the existence of a twice continuously
differentiable lensing potential).

The gravitational lensing of an infinitely thin relativistic
jet for elliptical lens galaxies described by
the models of a singular isothermal ellipsoid and an
ellipsoid with a core are considered in Section 1. The gravitational lensing of a jet by a spiral galaxy is
investigated in Section 2. In particular, we consider
the model of a disk and a softened halo located in an
isothermal dark matter halo for various values of their
parameters; the Kuzmin model of a disk (Kuzmin 1956) in an isothermal
halo; and the model of a disk and
bulge in an isothermal halo. Apart from investigating
the images of a relativistic jet that appear when it is
gravitationally lensed by galaxies described by both
one-component and two-components models of the
mass surface density distribution, the images emerging
in various models are also compared qualitatively
in the first two sections of the paper. For an adequate
comparison, it is necessary to fix the spatial
location of the relativistic jet whose choice is relatively arbitrary but the same for all models. In the succeeding
sections, when investigating the possibility of
the formation of ring structures as a result of the jet
lensing, we remove the requirement for its fixed spatial
location. The emerging images of the relativistic
counterjet for the models considered in Sections 1
and 2 are modeled in Section 3. The application of
our results to the source Â0218+357 is discussed in
Section 4. In Conclusions, we summarize our results
and discuss their possible applications.

\section*{GRAVITATIONAL LENS MODELS}

A detailed description of the theory of gravitational
lenses can be found in the book by Schneider
et al. (1999). Here, we provide only the basic definitions
needed for the subsequent understanding.

The lens equation maps the points of the lens plane
to the corresponding points of the source plane. For
all of the models considered here, the lens equation
will be written in dimensionless variables normalized
to the Einstein-Chwolson radius. The expression for
it is written out in the Appendix (Eq. A1).

The ratio of the flux density received by the observer
to the flux density that the observer would
receive in the absence of a lens is called the lens
magnification. The curve in the lens plane at each
point of which the lens magnification becomes infinite
is called a critical one. The curve in the source plane
that is the mapping of the critical curve is called a
caustic.

For the observed radio sources with compact images
and an extended ringlike structure, one of the
possible interpretations of their appearance is the situation
where they are lensed by a galaxy. In this
case, the compact images correspond to the image of
the source's central core and the extended structure
is associated with the image of the relativistic jet.
Generally, galaxies of various types can act as lenses,
but here we consider models for elliptical and spiral
galaxies. The expressions for the lensing potential and image magnification and the lens equation are
given in the Appendix, because they are cumbersome.
The lens equations for all models written out in analytic
form were solved numerically by the so-called
grid method where the region of presumed solutions
was scanned with a sufficiently small step. The grid
step was chosen to be $10^{-4}$ in both coordinates; the
residual of the solution obtained is better than $10^{-7}$.

The relativistic jet is represented as an infinitely
thin segment with a constant radiation intensity at
each of its points. The jet inclination $\alpha$ is measured
from the $x$ axis counterclockwise. The choice of
the jet inclination and the initial point of the jet are
arbitrary, but the tangential caustic crossing condition
is met in all of the models considered here. As
our analysis shows (see below), displacing the initial
point from the symmetry axis does not change fundamentally
the shape of the images; only the positions
of the initial points of the images and their number
change (when the initial point of the jet is located
outside the tangential caustic).

When modeling the jet images, we used the geometrical
optics approximation, which, as any approximation,
has its limitations. First, the light beams
coming from the lensed object to the observer may
turn out to be coherent, which will lead to interference
between these beams. Second, the magnification of
a point source as the latter approaches the caustic
tends to infinity. Despite these shortcomings, in an
overwhelming majority of the cases of gravitational
lensing, the geometrical optics approximation is justified
and the corrections introduced by wave optics
are significant only for very compact sources, for example,
extragalactic pulsars. This question was studied
in detail in the book by Schneider et al. (1999).
Here, we will only give an upper limit on the magnification
that emerges when the jet approaches the
tangential caustic from wave optics. In the model of an infinitely thin axisymmetric lens and the approximation
of a point source located near the astroid
at a cosmological distance, we can estimate
the maximum magnification defined by the expression
(Schneider et al. 1999)
\begin{equation}
\mu_{max}\sim(\frac{M}{M\odot})^{1/3}(\frac{\lambda}{10^6})^{-1/3},
\end{equation}
where $M$ is the lens mass and $\lambda$ is the wavelength
expressed in centimeters. For lenses with a mass
$M \sim 10^{10} M\odot$ in the radio band ($\lambda \sim 1$ cm), we obtain
$$\mu_{max}\sim10^{5-6}. $$

As we see, the magnification is not infinite but very
high, which introduces no changes in the results
obtained here.

\section{MODELS FOR ELLIPTICAL GRAVITATIONAL LENS GALAXIES}

As was noted in the Introduction, in most of the
observed gravitational lensing events with the appearance
of multiple images, the lenses are elliptical
galaxies that are described by the models of either
an elliptical density or an elliptical effective lensing
potential. An elliptical lens galaxy is best described
by the models of an isothermal singular ellipsoid and
an isothermal ellipsoid with a core. The shape of
the critical and caustic curves for these models was
studied in detail by Kormann et al. (1994).

Since the general expressions for the ellipsoid potential
are reduced only to elliptic integrals, we will
consider the special case of an ellipsoid - a spheroid
for which the expressions can be written via elementary
functions (Schmidt 1956). Let us introduce the
main notation and parameters used in the models
under consideration. Let $q_3$ be the axis ratio of the
spheroid. Its projection onto the plane perpendicular
to the line of sight to the source is then an ellipse with the axis ratio $q =\sqrt{q_3^2\cos^2(i)+\sin^2(i)}$, where $i$ is the
inclination of the spheroid. The angle $i = 0^o$ implies
that we see the spheroid edge-on; if, alternatively,
$i = 90^o$, then the projection of the spheroid onto the
plane is a circle.

The lens equation, the expressions for the lensing
potential, i.e., for the projection of the three dimensional
potential onto the lens plane, and the image
magnification are given in the Appendix (Eqs. A2
and A3).

The modeling results for a singular isothermal ellipsoid
and an ellipsoid with a core are presented in
Fig. 1. To be able to conveniently compare the cases
considered in the figure, we chose equal inclinations
of the jet $\alpha$ and its length of $60^o$ and $\sim 1.5$ Einstein-Chwolson radius, respectively. The dark gray and light-gray
lines indicate the caustic and critical
curves, respectively; the black line indicates the relativistic
jet projected onto the lens plane. The color of the jet images correspond to different
magnifications. Purple color is used to denote regions where the absolute value of the magnification is less than 1, blue is for magnifications changing from 1 to 3, green - from 3 to 7, yellow - from 7 to 10, and red is for magnifications greater than 10.  The upper
and lower case letters denote some of the reference
points on the jet and the corresponding points on the
images, respectively.

The radiation intensity along the jet is assumed to
be constant in all of the models under consideration,
which, of course, disagrees with available observations.
However, if the magnification at each point of
the jet is known, then the intensity distribution can
be derived from its images by specifying an arbitrary
intensity distribution along the jet.

Consider the model of an isothermal ellipsoid with
a core $s = 0.3$ and spheroid axis ratios typical of elliptical galaxies, $q_3 = 0.3$ (Fig. 1a) and $q_3 = 0.6$
(Fig. 1b). We see that as the parameter $q_3$ increases,
the size of the critical and caustic curves decreases
and the jet image slightly contracts, while its shape
is retained. On the whole, the jet image consists of
individual fragments separated by the critical curves.
For case (a), when moving along the jet from quadrant
III upward into quadrant I until the crossing
of the tangential caustic in the shape of an astroid,
one image is observed (point \emph{a}). When crossing the tangential caustic, two new images appear in
quadrant II: one moves from the outer critical curve
upward into quadrant I and the other moves from the
outer critical curve downward toward the inner critical
curve (points \emph{b}). Subsequently, when crossing the
radial critical curve in the shape of an ellipse, two additional
images appear in quadrant I: one moves from
the inner critical curve rightward toward the outer
critical curve in quadrant IV and the other moves from
the inner critical curve in quadrant I leftward toward
the same curve in quadrant III (points \emph{c}, \emph{d}). Thus,
five jet images are formed in the region within both
caustics. The inverse process, i.e., the merging of the
images, takes place when moving upward along the
jet and crossing of the caustics. When going outside
the radial and tangential caustics, respectively,
three (points \emph{e}) and one (point \emph{f}) images remain. Four
regions of significant magnification indicated by the
red color appear for the specified jet configuration.
\begin{figure*}
\begin{center}
\includegraphics[width=0.42\textwidth]{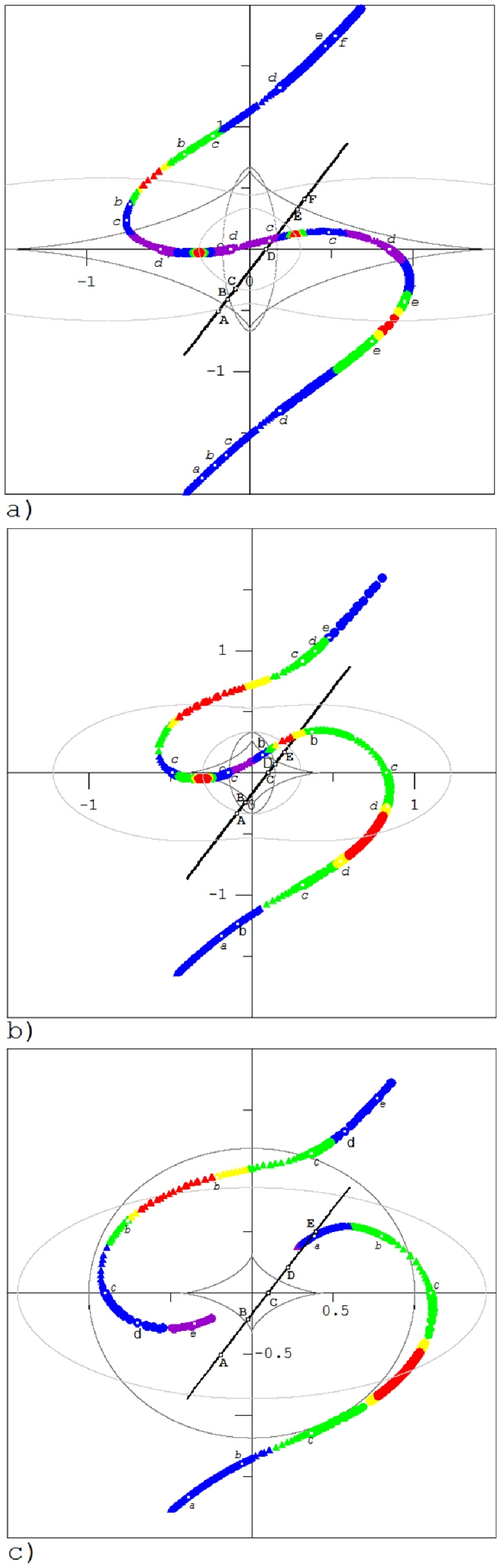}
\caption{\rm Results of modeling a relativistic jet for the models of a singular ellipsoid and an ellipsoid with a core: (a) $q = 0.3$ and
$s = 0.3$, (b) $q = 0.6$ and $s = 0.3$, (c) $q = 0.6$ and $s = 0.0$. The dark gray and light-gray lines indicate the caustic and critical
curves, respectively; the extended jet is represented by the black line; the jet images as a function of magnification are indicated by different colors. Purple corresponds to the case when $|M|<1$, blue - $1<|M|<3$, green - $3<|M|<7$, yellow - $7<|M|<10$ and red - $|M|>10$.}
\label{NIE&SIE}
\vfill
\end{center}
\end{figure*}

In contrast to Fig.~1a, for case (b), the radial caustic
is initially crossed and two images appear in quadrant
I on the inner critical curve: one moves rightward
toward the outer critical curve in quadrant IV and the
other moves leftward into quadrant III from the inner
critical curve in quadrant I (points \emph{b}). When crossing
the tangential caustic, two more images appear in
quadrant II: one moves from the outer critical curve
upward into quadrant I and the other moves from
the outer critical curve downward toward the inner
critical curve (points \emph{c}). Subsequently, just as in
case (a), the images merge together as the jet moves
outside the regions of the caustic curves.

For comparison, consider the model of a singular
isothermal ellipsoid (Fig.~1c). The presence of a
central singularity gives rise to a region in the source
plane outside the caustic where multiple images also
exist. The curve that bounds this region is called a
cut (Kormann et al. 1994). When crossing this curve,
an ``infinitely faint'' image is formed at the coordinate
origin of the lens plane that moves from quadrant I
toward the critical curve in quadrant IV (point \emph{a} in
Fig.~1c). When crossing the astroid caustic, two
images appear in quadrant II: one moves from the
critical curve upward into quadrant I and the other
moves from the critical curve downward toward the lens center through quadrant III. Thus, four jet images
(points \emph{b}, \emph{c}) are formed in the region within the
caustic and the cut, with two images located within
the critical curve having the same negative parity.
When emerging from the region of the astroid caustic,
the two images merge together in quadrant IV.
Two images (points \emph{d}, \emph{e}) remain in the region outside
the caustic but inside the cut. One image remains
outside the latter.

As we see from Fig.~1, the images for the cases
considered are quite similar, but, in contrast to an
ellipsoid with a core, only two regions of high magnification
(red arcs, Fig.~1c) appear in the case of a
singular isothermal ellipsoid.

Let us investigate the influence of the jet inclination
$\alpha$ on the picture of its gravitational lensing. Figure
2 shows the modeling results for an isothermal ellipsoid
with a core equal to $0.3$ and a spheroid axis ratio
$q_3 = 0.6$ for two jet inclinations, $\alpha = 45^o$ (Fig.~2a)
and $\alpha = 90^o$ (Fig.~2b). Our studies showed that for
fixed parameters of the ellipsoid at jet inclinations that
do not coincide with the coordinate axes (symmetry
axes), the shape of the jet image is a double loop and
does not change significantly with $\alpha$ (Figs.~1b, 2a).
However, at an inclination, for example, of $90^o$, the
shape of the jet images changes; as a result, a loop
is observed in the direction of the unlensed jet and a
ring about $0.3$ Einstein-Chwolson radius in diameter
is observed in the opposite (relative to the $y$ axis)
region. The two images that form this ring appear in
quadrant II and merge together in quadrant III when
crossing the ellipse caustic.

At fixed spheroid axis ratio $q_3$, a change in inclination
$i$ causes a change in the axis ratio of the
spheroid projection $q$. In the case of a singular ellipsoid,
the caustic is reduced with increasing inclination
and degenerates into a point at $i = 90^o$, implying
the appearance of two jet images without bright
arcs. The behavior of the caustics in the case of an
isothermal ellipsoid with a core was studied in detail
by Kormann et al. (1994). They showed that the
caustics have different topologies, depending on the
axis ratio of the ellipse projected onto the lens plane
and the core radius. Thus, the shape and size of the
caustics turn out to be important for the appearance
of high-magnification regions, because the number of
emerging images for a jet of fixed length is determined
by its location relative to the caustics.
\begin{figure*}
\begin{center}
\includegraphics[width=0.99\textwidth]{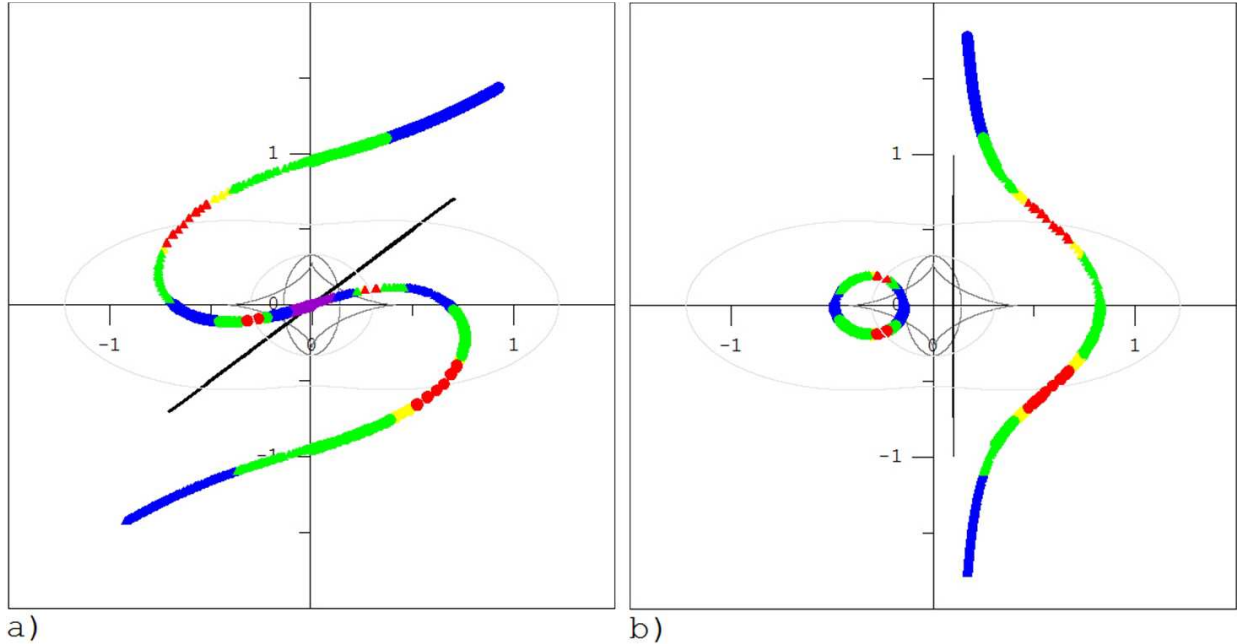}
\caption{\rm Results of modeling a relativistic jet for the model of a singular ellipsoid with a core with parameters $q = 0.6$ and $s = 0.3$
at different jet inclinations to the $x$ axis: (a) $\alpha = 45^o$ and the jet passes through the coordinate origin; (b) $\alpha = 90^o$ and the jet
passes through the point (0.1, 0). The designations of the critical and caustic curves, the jet, and its images are the same as
those in Fig.~1.}
\label{NIE_ang}
\vfill

\end{center}
\end{figure*}
\section{MODELS FOR SPIRAL GRAVITATIONAL LENS GALAXIES}
\subsection{Model I}
When modeling the mass surface density distribution
of a spiral galaxy, we will follow the approach proposed by Keeton and Kochanek (1998). We will
begin our modeling with the simplest case. Consider
a singular disk truncated at a characteristic distance
$a_d$ from the center and placed in an isothermal halo
with a characteristic size $a_h$. We will call this model
``model I''.

The rotation curves of spiral galaxies are the main
tool for determining the mass distribution in these
galaxies. According to observational data, depending
on the spiral galaxy type, the rotation velocity peak for
Sa, Sb, and Sc is reached approximately at 300, 220,
and $175 ~km~s^{-1}$, respectively, and the rotation curve
subsequently becomes flat (Sofue and Rubin 2001).
No significant deviations between the rotation curves
of high-redshift spiral galaxies and those of nearby
spirals have been found (see, e.g., Kelson et al. 2000).

To take into account the contribution of dark matter
to the rotation curve at distances $R < a_d$, we will
introduce the factor $f_d$. The equality of $f_d$ to one
implies that the rotation curve is entirely determined
by the visible disk component. It should be noted
that analysis of the disk surface densities obtained
by different methods showed that the disk does not
contain a large amount of dark matter (Holmberg
and Flynn 2004). Here, this factor was taken to be
$f_d = 0.85$ (Sackett 1997).

The lensing potential for model I, the lens equation,
and the expressions for the image magnification
are given in the Appendix.

To model the jet images produced by gravitational
lensing for the potential of model I, we will fix the following model parameters:
\medskip

\begin{tabular}{c|c}
$f_d$   & $0.85$\\
\hline
$a_h$   & $0.8 a_d$ \\
\hline
$q_{3h}$& 0.8\\
\end{tabular}

\medskip

The choice of the parameters is dictated by two
main criteria: the halo shape is nearly spherical
(Dehnen and Binney 1998) and the rotation curve
of a spiral galaxy is flat. At fixed jet parameters,
different caustic and cut geometries and, as a result,
different image geometries are observed depending
on $q_d$ and $a_d$. In Fig. 3, except for the case of
$q_d = 0.01$ and $a_d = 100.0$ (Fig.~3b) that corresponds
to a ``disk geometry'', the relativistic jet crosses the
caustic. As the caustic curve is approached, the
images with different parities approach each other,
merging together when crossing the caustic. This is
characterized by the appearance of bright extended
arcs. In the case of a disk geometry, the caustic ``pierces'' the cut and pairwise symmetric images
appear; the central images are faint, while the brighter
images are separated by a considerable distance (of
the order of four Einstein-Chwolson radii). Note
that as yet no such case has been encountered in
observations (see, e.g., the CASTLES catalog\footnote{http://www.cfa.harvard.edu/castles/noimages.html}),
which probably suggests that the halo should be
taken into account for spiral lens galaxies.
\begin{figure*}
\begin{center}
\includegraphics[width=0.42\textwidth]{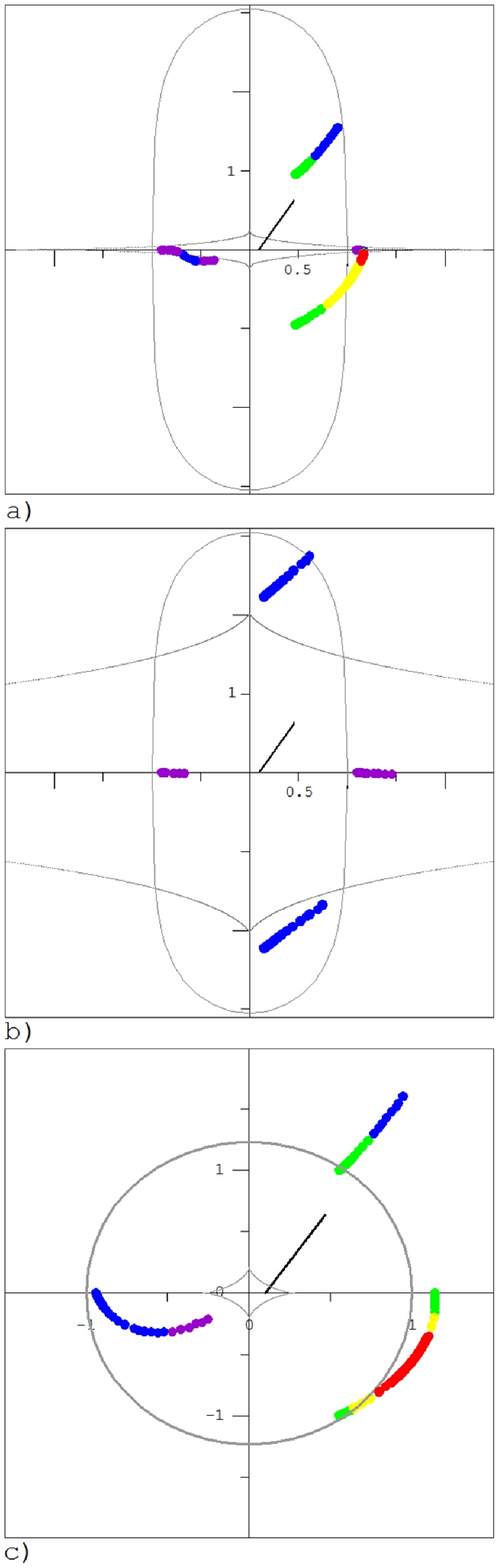}
\caption{\rm Results of modeling a relativistic jet for model I at
$i = 0^o$: (a) $q_{3d} = 0.01$ and $a_d = 1.0$; (b) $q_{3d} = 0.01$ and
$a_d = 100.0$; (c) $q_{3d} = 0.5$ and $ad = 1.0$. The designations
are the same as those in Fig.~1.}
\label{mod1}
\vfill

\end{center}
\end{figure*}
\begin{figure*}
\begin{center}
\includegraphics[width=0.42\textwidth]{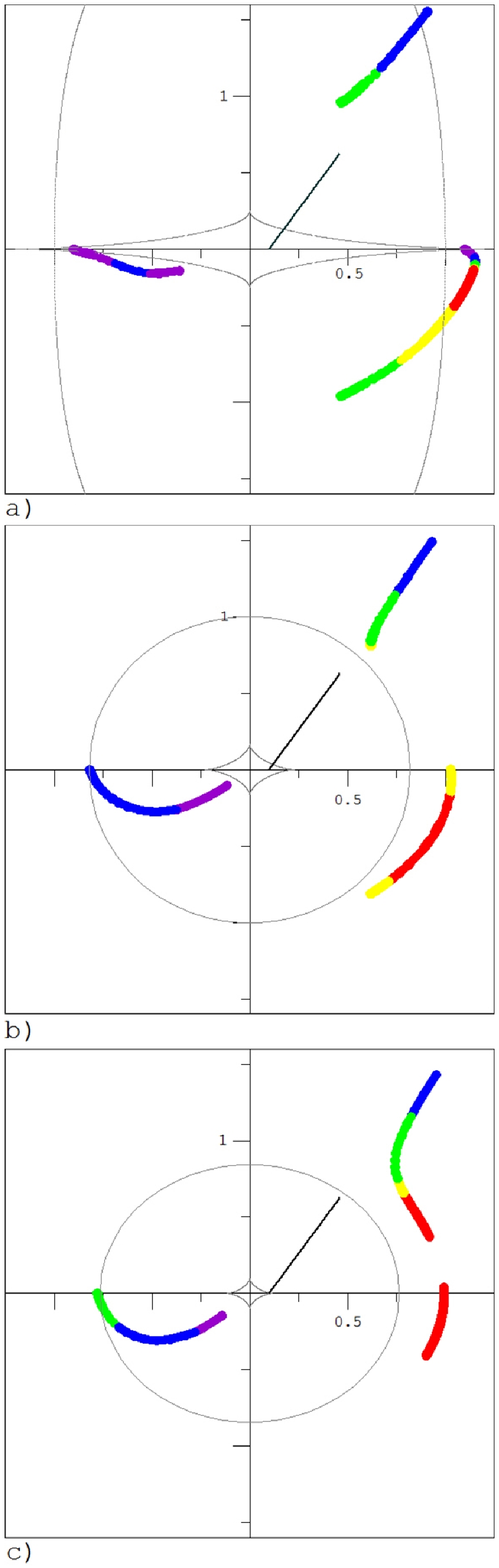}
\caption{\rm Results of modeling a relativistic jet at various
inclinations $i$ for model I with $q_{3d} = 0.05$ and $a_d = 1.0$;
(a) $i = 0^o$; (b) $i = 30^o$, (c) $i = 45^o$. The designations are
the same as those in Fig.~1.}
\label{i}
\vfill

\end{center}
\end{figure*}
\begin{figure*}
\newpage
\begin{center}
\includegraphics[width=0.9\textwidth]{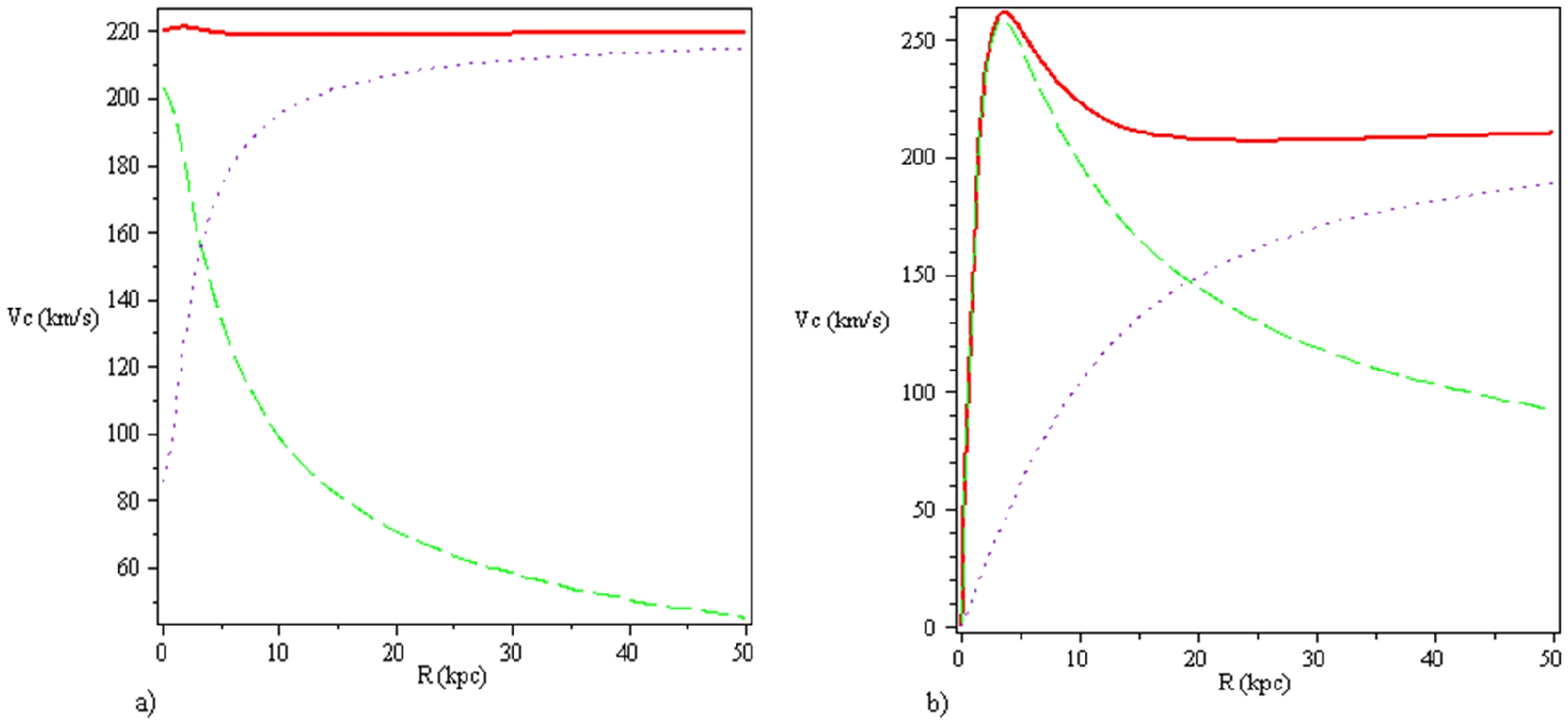}
\caption{\rm Rotation curves for models I(a) and II (b) calculated for $q_d = 0.05$ and $a_d = 1.0$. The red solid, green dashed, and purple dotted lines
indicate the total rotation curve, the disk contribution, and the halo contribution, respectively.}
\label{rcurves}
\vfill

\end{center}
\pagebreak
\end{figure*}
Based on model I, let us investigate the influence of
the inclination of the spheroids that describe the disk and halo. For $q_{3d} = 0.05$ and $a_d = 1.0$, Fig.~4 shows
the jet lensing for coaxial spheroids and inclinations $i$
of $0^o$, $30^o$, and $45^o$. Note that the choice of the above
values for $q_{3d}$ and $a_d$ is dictated by the Milky Way
parameters (see, e.g., Grimm et al. 2002). The mean
axis ratio of spiral galaxies is $\sim 0.1-0.2$ (de Grijs
and van der Kruit 1996). However, a change in
$q_{3d}$ from 0.05 to 0.1 has no significant effect on the
shape of the jet images at other identical model and
jet parameters. We see from the figure that as the
inclination increases, the caustic (astroid) is reduced
and the shape of the images changes. As with the
model of a singular isothermal ellipsoid, the caustic
degenerates into a point when $90^o$ is reached.

The expression for the galactic rotation curve in
model I is given in the Appendix (Eq. A21), while the
shape of the curve itself with disk parameters $q_{3d} =
0.05$ and $a_d = 1.0$ is shown in Fig.~5a. We see from the
figure that the rotation curve in this model at small $R$,
i.e., near the galactic center, does not correspond to
the observed rotation curves of spiral galaxies (Sofue
and Rubin 2001).

To compare the modeling results obtained for
model I with the complicated models considered
below, we will use the approach described in Section 1
of this paper. For the disk parameters considered
above, we will choose such a length of the jet that
it crosses both the caustic and the cut twice (Fig.~6a).
A jet with $\alpha = 60^0$ crosses the $x$ axis at a point of
$0.1$ Einstein-Chwolson radius. All designations in
Fig. 6 are the same as those in Fig.~1. The shape
of the images, except for some features, and their
behavior for model I are similar to those in the case of
a singular isothermal ellipsoid, but the critical curves
differ significantly (Fig.~6a).

\subsection{Model II}

Consider the more realistic model of a spiral galaxy
that is a galactic disk approximated by a Kuzmin disk
(Eqs. A23 and A24) embedded in a halo modeled
by a nonsingular isothermal ellipsoid. We will call
this model ``model II''. The lensing potential for this
model, the lens equation, and the expression for the
image magnification are given in the Appendix. The
model under consideration is described by the following
parameters: the characteristic disk, $a_d$, and
halo, $a_h$, sizes, the axis ratios in the disk $q_{3d}$ and halo
$q_{3h}$, and the disk mass $m_d$. These parameters were
determined from the following conditions: (i) the rotation
curve characteristic of spirals, (ii) the specified
Galactic rotation velocity of the Sun $V_0(8.0 ~kpc) =
235~km~s^{-1}$ (Reid and Brunthaler 2004), and (iii) the
local disk surface density in the solar neighborhood
$\Sigma_{\odot}=75\pm25~M_{\odot}~pc^{-2}$. They are given below:

\medskip
\begin{tabular}{|c||c|c|}
\hline
    &$q_{3d}$   &$a_h$\\ \cline{2-3}

$a_{d}=1.0\,$   &$0.01$ &$4.63$    \\ \cline{2-3}

$m_{d}=1.0\cdot 10^{11}M\odot$    &$0.05$ &$4.36$    \\ \cline{2-3}

$q_{3h}=0.8$    &$0.10$  &$4.07$\\ \cline{2-3}
    &$0.15$  &$3.83$ \\
\hline
\end{tabular}

\medskip

Figure~5b for $q_{3d} = 0.05$ and $a_d = 1.0$ shows
the rotation curve for model II in comparison with
model I.

It should be noted that a Kuzmin disk approximates
better the galactic disk than a nonsingular
isothermal ellipsoid, because, in this case, the rotation
curve coincides in shape with the rotation curves
observed for spiral galaxies.

Radial and tangential caustics appear for typical
$q_{3d}$ of spiral galaxies in model II; in contrast to
the previously considered models of the mass distribution
in the lens, part of the astroid lies outside
the radial caustic. The caustic curves of this type
are called ``naked cusps''. Figure~6b presents the jet
modeling results in model II for the same galactic
disk parameters as those in model I. The shape of
the emerging images and their behavior are similar to
those in the case of a nonsingular isothermal ellipsoid,
but the critical curves are different. The number of
emerging regions of enhanced brightness is twice that
in model I. This corresponds to four crossings of the
critical curves, although the extent of these regions is
considerably smaller.

\subsection{Model III}

\begin{figure*}
\begin{center}
\includegraphics[width=0.42\textwidth]{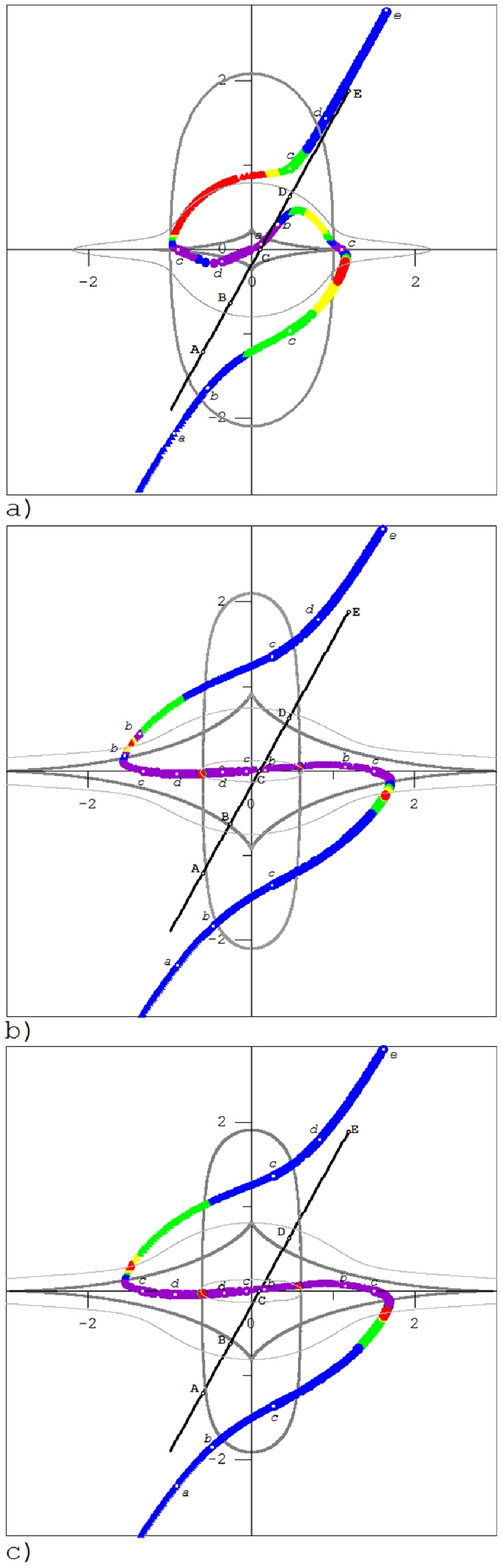}
\caption{\rm Comparison of the results of modeling a relativistic
jet for models I, II, and III. The jet geometry ($\alpha =
60^o$) and model parameters ($q_{3d} = 0.05$ and $a_d = 1.0$) are
fixed; (a) for model I, (b) for model II, and (c) for model III.
The designations of the critical and caustic curves, the jet,
and its images are the same as those in Fig.~1.}
\label{all}
\vfill

\end{center}
\end{figure*}

\medskip


Let us investigate the question of whether the
bulge in the spiral galaxy structure should be taken
into account when considering the gravitational lensing
of a jet by adding a bulge in the shape of a Kuzmin
disk to model II (model III, Eq. A30).

We determine the model parameters from the condition
that the rotation curves for models II and III
coincide for given $q_{3d}$. In the model, we chose the
bulge axis ratio $q_{3b} = 0.6$ and the characteristic core
radius $a_b = 0.8$ (Dehnen and Binney 1998). The fixed
parameters of model II are given below:
\medskip

\begin{tabular}{|c||c|c|c|}
 \hline
$a_{d}=1.0\,$    &$q_{3d}$   &$m_b, 10^{10}M\odot$   &$a_h$\\ \cline{2-4}

$m_{d}=8.8\cdot 10^{10}M\odot$   &$0.01$  &$1.37$ &$4.68$    \\ \cline{2-4}

$q_{3h}=0.8$    &$0.05$    &$1.36$ &$4.42$    \\ \cline{2-4}

$q_{3b}=0.6$    &$0.10$  &$1.34$  &$4.13$\\ \cline{2-4}
$a_b=0.8$    &$0.15$  &$1.32$  &$3.88$ \\

\hline
\end{tabular}

\medskip

Figure~6c presents the jet modeling results for
model III in comparison with models I and II (Figs.~6a,
6b). The shape of the emerging images, their behavior,
and the critical and caustic curves in models II
and III are essentially identical, suggesting that the
bulge introduces no significant change in the picture
of gravitational lensing of the jet for the chosen model
parameters.

Note that all three galactic components are assumed
to be coaxial, although, according to observations,
this condition is not met for the bulge. Our
studies for noncoaxial components showed that the
modeling results obtained in this case differ only
slightly from those for coaxial ones. For example, if
the angle between the disk and halo symmetry axes
is $45^o$ and between the disk and bulge in model III is
$20^o$, then the maximum displacement of the jet image
compared to that for coaxial galactic components in models II and III is only $0.05$ Einstein-Chwolson
radius. For model I, Fig.~7 shows the jet images for
the coaxial and noncoaxial cases. We see from this
figure that the shape of the images does not change
significantly, but only a slight displacement, whose
maximum value does not exceed $\sim0.1$ Einstein-Chwolson radius, takes place. It should be noted
that when the jet crosses the caustic curve near the
cusp, the maximum displacement can reach $\sim0.2$ Einstein-Chwolson radius.


\begin{figure*}
\begin{center}
\includegraphics[width=0.7\textwidth]{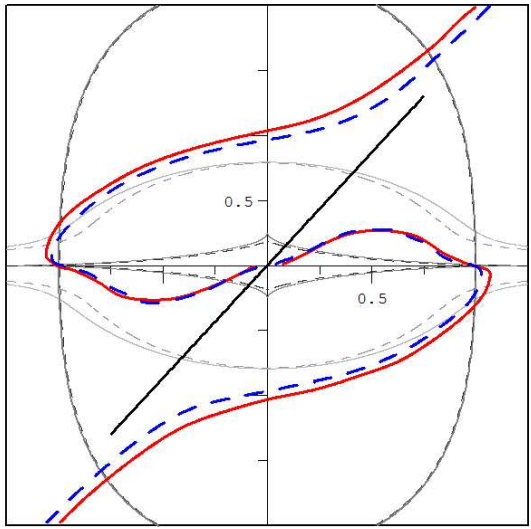}
\caption{\rm Results of modeling the images of a relativistic jet
for model I in the case of coaxial components (red solid
thick line) and noncoaxial components (blue dashed thick
line), when the angle between the disk and halo symmetry
axes is $45^o$. The corresponding caustic and critical curves
are indicated by the thin solid and dashed lines. The jet
inclination (indicated by the black color) is $\alpha = 60^o$; the
model parameters are $q_{3d} = 0.05$ and $a_d = 1.0$.}
\label{nesoosnyj}
\vfill

\end{center}
\end{figure*}

 \section{THE COUNTERJET}

For a given mass surface density distribution in
the lens galaxy, the observed shape and relative positions
of the images in the plane of the sky are
determined by the position of the initial point and
the jet length and inclination relative to the caustic
curves. Different configurations of jet images both
in their number and parity and in magnification at
each image point and the presence of bright extended
arcs can be obtained by changing these parameters
for a specific model of the lens galaxy. The gravitational
lensing of relativistic jets can, in principle,
allow the counterjet that cannot be observed in the
absence of gravitational lensing due to the geometry
to be observed. This, in turn, can be a test for the
generation models of relativistic jets when discussing the question of whether such a counterjet is present
or absent. Based on the lens models considered, we
investigated the question of how the counterjet would
manifest itself during gravitational lensing. It turned
out that a great variety of lensing pictures, including
extended arcs and almost circumferences (under the
caustic curve crossing condition), could be formed
by changing its length and direction. For ringlike jet
images consisting of bright extended arcs to appear,
the jet must cross the tangential caustic almost along
the tangent to its cusps.

A characteristic example of the appearance of such
a ringlike structure is shown in Fig.~8 for model I.
The initial point of the jet marked by the asterisk
lies outside the tangential caustic and is inclined at
the angle $\alpha = 60^o$. to the $x$ axis (as in the previously
considered cases); the remaining parameters of the
chosen model are $q_{3d} = 0.05$, $a_d = 1.0$, and $i = 30^o$.

\begin{figure*}
\begin{center}
\includegraphics[width=0.5\textwidth]{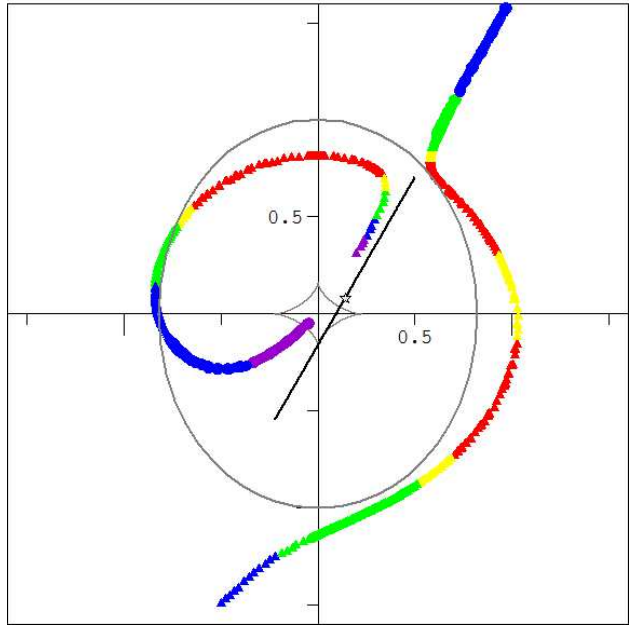}
\caption{\rm Results of modeling a relativistic jet and counterjet
for model I at $i = 30^o$, $q_{3d} = 0.05$, and $a_d = 1.0$.
The initial point of the jet designated by the asterisk is
located outside the astroid and has coordinates $(0.14,
0.08)$ in the lens plane. The designations are the same
as those in Fig.~1.}
\label{counterjet}

\vfill

\end{center}
\end{figure*}

\section{B0218+357}

Based on the results of the above modeling, we
can assume that some of the observed gravitationally
lensed systems with large-scale rings can be explained
by the lensing of their jets. It may well be that
the source B0218+357, in which two compact core
images and an extended ring structure are observed
in the radio band, is such an object. In particular,
5-GHz MERLIN observations of this source showed that the ring structure consists of two arcs each of
which, in turn, breaks up into several separate regions
of enhanced brightness (Biggs et al. 2001).

An attempt to model the lens mass distribution
in the system B0218+357 was made by Wucknitz
et al. (2004). As the model that described most
adequately the observed picture, the authors used a
singular isothermal elliptical power-law lensing potential
specified by the formula
\begin{equation}
\psi(x,y)={\theta_0}\sqrt{\frac{x^2}{(1+\epsilon)^2}+\frac{y^2}{(1-\epsilon)^2}},
\end{equation}
where $\theta_0$ is the Einstein-Chwolson angle and $\epsilon$ is the
ellipticity of the distribution.

According to VLA data, the large-scale jet is inclined
with respect to the line connecting the
source's two images A and B at the angle $\varpi$ that
has not be determined accurately, but, judging by its
VLA images at 8.4 GHz (Fig. 4 in Biggs et al. 2003),
it is approximately equal to $75^o$; the image intensity
ratio is $F_A/F_B\simeq2.2-3.9$, depending on frequency
(see Table 3 in Mittal et al. 2006). If, based on the
model proposed by Wucknitz et al. (2004), we direct
the jet precisely at this angle, then its images will be
dim, slightly curved lines that are hard to associate
with a ring. The following question arises: Can
a ringlike structure still be obtained in this model
and what is needed for this? As was pointed out
above, bright arclike structures appear when the jet
approaches the caustics (or when it crosses them). In the case of Wucknitz's model, a counterjet that
will cross the tangential caustic (astroid) is needed
to obtain such a structure. However, in this case,
the jet itself must be directed at an angle $\varpi\simeq30^o$
and the counterjet length must be comparable to
the Einstein-Chwolson radius, which is actually not
observed.

\begin{figure*}
\begin{center}
\includegraphics[width=0.8\textwidth]{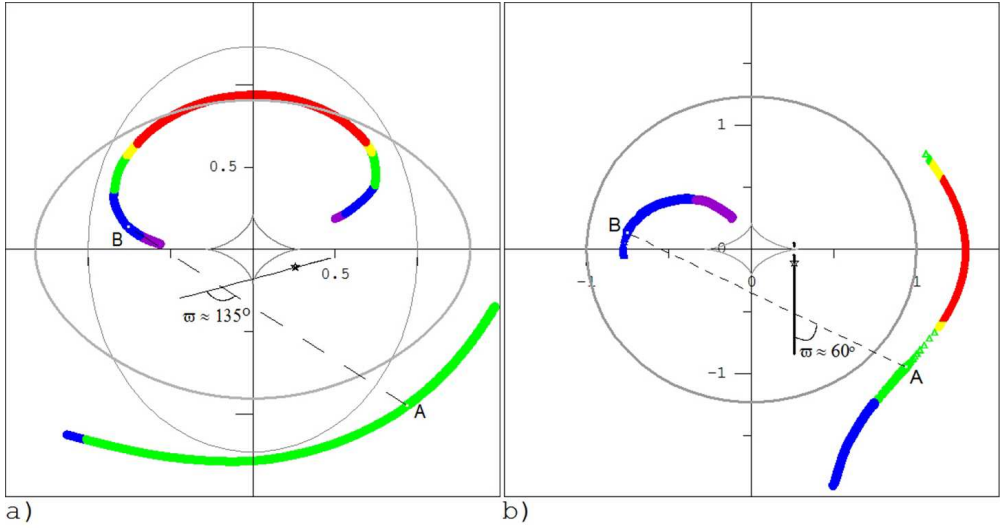}
\caption{\rm Results of modeling a relativistic jet for model I at $q_d = 0.5$ and $a_d = 1.0$ for the gravitational lensing of B0218+357.
The initial point of the jet that corresponds to the position of the source's compact core has coordinates $(0.26, -0.11)$ in the
lens plane. Panel (a) illustrates the situation where the jet is directed at an angle $\varpi\simeq135^o$ with respect to the line connecting
images A and B and crosses the lower edge of the tangential caustic; in panel (b), the jet is directed at an angle $\varpi\simeq60^o$ and
has a counterjet with a length approximately equal to one fifth of the Einstein-Chwolson radius that crosses the right end of
the astroid. The designations are the same as those in Figs.~1 and 6.}
\label{b0218}
\vfill

\end{center}
\end{figure*}

An interesting question is whether the parameters
and the position of the initial point of the jet
can be chosen within the framework of models I-III
considered here to explain the observed geometry of
B0218+357, including its large-scale structure consisting
of two extended arcs. Our study showed that
different sets of parameters that allow the observed
geometry of the system and the intensity ratio of the
emerging images to be obtained in the models considered
(see above). However, obtaining bright extended
arcs is very nontrivial and is possible by no means in
all cases. As one of such realizations, we can choose
model I with a characteristic disk size $a_d = 1.0$ and
an axis ratio of its projected surface density ellipse
$q_d = 0.5$. If we place the initial point of the jet at
the point with coordinates $(0.26,-0.11)$, then the
distance between the images of the compact core is
$\simeq 335 ~mas$ and the magnification ratio of its images
is $\simeq 3.3$, in good agreement with observational data
(Biggs et al. 1999; Patnaik et al. 1993). Bright semiring
structures (arcs) in this model appear at an angle
$\varpi\simeq135^o$ and no counterjet is required in this case (Fig.~9a). It should be noted that the angle between
the jet and the line connecting the images for the chosen parameters differs significantly from the observed
one. If we direct the jet at an angle $\varpi\simeq60^o$ for
the same model parameters and add the counterjet,
then we will obtain the caustic crossing and bright
arcs, as is illustrated in Fig.~9b. In this case, the
angle $\varpi$ is fairly close to the observed one and the
counterjet length accounts for less than one fifth of
the Einstein-Chwolson radius. In addition, multifrequency
VLBI observations at 1.65 GHz revealed
a component in image A in the direction opposite
to that of the component associated on small scales
with the jet relative to the compact core that can be
associated with the counterjet (Mittal et al. 2006).
At the same time, this component is not observed in
image B, which is explained in terms of the models
considered by the fact that the fainter image B and
the corresponding image of the jet are spatially more
``compressed'' than the brighter image A (see, e.g.,
Fig.~9b).

\medskip

\section{CONCLUSIONS}

We modeled the images of relativistic jets from
extragalactic sources produced by gravitational lensing
by galaxies of various types. To describe the
surface density distribution for elliptical lens galaxies,
we used the models of a singular isothermal ellipsoid
and an ellipsoid with a core; for spiral lens galaxies,
we considered the model of a disk and a softened
disk embedded by a singular isothermal dark matter halo,
the Kuzmin model of a disk in an isothermal halo,
and the model of a disk and a bulge in an isothermal
halo. The critical and caustic curves and the relativistic
jet images for the three multicomponent models
of spiral galaxies presented here were compared
with one another. For the chosen parameters, the
shape of the emerging images, their behavior, and
the critical and caustic curves are almost identical for
models II and III. This suggests that including a low mass
bulge similar to the Milky Way bulge introduces
no significant changes in the picture of gravitational
lensing of the jet.

We showed that the observed large-scale ring
structures could be produced by the gravitational
lensing of relativistic jets by galaxies for certain
relative positions of the jet and the caustic curves.
In particular, for extended bright arcs to appear in
the models considered here, the jet must cross twice
the tangential caustic almost along the tangent to its
cusps.

For the gravitationally lensed system Â0218+357,
we compared the model used previously to determine
the Hubble constant (Wucknitz et al. 2004) with the
models of the mass surface density distribution in a
spiral lens galaxy proposed here. For example, in the
model with a singular elliptical power-law potential mentioned above, it is highly problematic to obtain
jet images in the shape of a ring structure for the
observed (outside the lensing region) direction of the
large-scale relativistic jet relative to the compact core
images. The models being discussed here that allow
for the gravitational lensing of a large-scale jet give an
image of the system Â0218+357 that, on the whole,
closer to the observed one.

In view of its ``isolated'' spatial location and the
measured time delay between its images, the system
Â0218+357 is well suitable for an independent
determination of the Hubble constant. However,
apart from the time delay, this requires an accurate
knowledge of the system's geometry, in particular,
the relative positions of the lens galaxy and the
emission source. Deep optical HST observations
of Â0218+357 aimed at determining the position of
the lens galaxy, along with the use of the model by
Wucknitz et al. (2004), allowed the Hubble constant
to be estimated, which changes depending on the
assumptions of the authors regarding the spiral arms
of the lens galaxy (York et al. 2005). Since the distance
between the images of the compact core for the
source under consideration in the radio band is fairly
small, $\simeq335 ~mas$, (it is slightly smaller in the optical
band) and since the lens is a spiral galaxy, optical
observations are very complicated. Submillimeter observations
are preferred for determining the position of
the lens galaxy due to the emission of cold dust in its
spiral arms. However, the resolution of present-day observatories in this band is not yet sufficient, which
makes the direct determination of accurate relative
positions of the lens and the source in the immediate
future problematic. Therefore, to reconstruct the geometry
of a gravitationally lensed system, modeling
remains topical. As was noted above, the Hubble
constant $H$ determined in the model also depends
on the relative positions of the lens and the source.
For example, Wucknitz et al. (2004) provided $H\simeq
78~km~s^{-1}Mpc^{-1}$, while York et al. (2005) gave $70$
and $61 ~km~s^{-1}Mpc^{-1}$, depending on the method of
determining the lens position. A significant difference
of the relative spatial positions of the lens and
the source in the models proposed here from those
obtained by other authors and between themselves
leads to a considerable variety in the estimates of the
Hubble constant, $H \simeq 35 - 90$ $km ~s^{-1} Mpc^{-1}$.

As another possible astrophysical application of
our results, we will note the possibility of measuring
the propagation velocity of jet knots during the observation
of its gravitationally lensed images. This
possibility follows from the following considerations.
In the models considered here, first, the lensed jet
image located outside the Einstein-Chwolson radius
is always ``extended'' compared to the jet projection
onto the lens plane. Second, when the jet crosses the caustic, the sizes of the emerging bright arc exceed
the initial sizes of the jet projection by many times.
These peculiarities make it possible to calculate the
jet propagation velocity by measuring the displacements
of individual bright knots in the images with
time. However, such studies are a separate independent
problem and are beyond the scope of this paper.

\section{ACKNOWLEDGMENTS}

This work was supported by the ``Origin, Structure,
and Evolution of Objects in the Universe'' Program
of the Presidium of the Russian Academy of
Sciences, the Program for Support of Leading Scientific
Schools (project no. NSh-5069.2010.2), and
the State contract P1336. We are also grateful to the
referees for a careful reading of the paper and helpful
remarks.

\bigskip

\section{APPENDIX}

All of the dimensional variables, including the
variables $x$ and $y$ and the parameter $s$, were normalized
to the Einstein-Chwolson radius. The
Einstein-Chwolson radius is a characteristic lensing
scale in the lens plane. For axisymmetric lens models,
it is defined by the formula
\begin{equation}
\label{thetaE} \xi_0=\sqrt{\frac{4Gm}{c^2}\frac{D_dD_{ds}}{D_s}},\tag{A1}
\end{equation}
 where $m$ is the lens mass, $D_d$ and $D_s$ are the distances from the observer to the lens and the source, reapectively, and $D_{ds} = D_s-D_d$ is the distance between the source and the lens.

The lens equation is
\begin{equation} \label{lenseq}
     \left\{
\begin{array}{rcl}
X&=& x-\Phi_{x} \\ \tag{A2}
Y&=& y-\Phi_{y} \\
\end{array}
\right.
\end{equation}
where $(X, Y)$ specify the jet points, $\Phi_x$ and $\Phi_y$ are the first derivatives of the lensing potential $\Phi$.

The image magnification is
\begin{equation} \label{magn} M^{-1}=1-\Delta\Phi+\Phi_{xx}\Phi_{yy}-\Phi_{xy}^2, \tag{A3}
\end{equation}
 where $\Phi_{xx}, \Phi_{yy}, \Phi_{xy}$ are the second derivatives of the lensing potential $\Phi$.

 For a multicomponent lens model, the lensing
potential is the sum of the potentials of its individual
components, $\Phi = \sum \phi_i$. Accordingly, the derivative
of the potential $\Phi$ - is the sum of the derivatives of the potentials $\phi_i$:
\begin{equation} \label{sum} \Phi_{x(y)}=\sum \phi_{ix(y)}. \tag{A4} \end{equation}

\begin{center} \textbf{A Singular Isothermal Ellipsoid and an Ellipsoid
with a Core}\end{center}

The lensing potential, i.e. the projection of the three-dimensional potential onto the lens plane, is defined by the formula
\begin{equation} \label{NIE}
\phi(s, q)=x\alpha_x+y\alpha_y-\frac{1}{2}bs\ln[(\psi+s)^2+(1-q^2)x^2], \tag{A5}
\end{equation}
\begin{equation} \begin{split}\alpha_x&=\frac{b}{\sqrt{1-q^2}}\arctan\frac{\sqrt{1-q^2}x}{\psi+s},\\
\alpha_y&=\frac{b}{\sqrt{1-q^2}}arth\frac{\sqrt{1-q^2}y}{\psi+q^2s},  \end{split} \tag{A6}
\end{equation}
where $e=\sqrt{1-q^2}$, $b = e/ \arcsin(e)$, $\psi=\sqrt{q^2(s^2+x^2)+y^2}$.

The first derivatives of the potential $\phi(s, q)$ are

\begin{equation} \label{firstder}
    \begin{split}
&
\phi(s, q)_x=\alpha_x+bx\frac{\psi-\frac{q^2(x^2+s^2)}{\psi}}{Z_1}-b\frac{\frac{q^2xy^2}{\psi}}{Z_2}, \\
&
\phi(s,q)_y=\alpha_y-b\frac{sy+\frac{y(x^2+s^2)}{\psi}}{Z_1}+by\frac{\psi+q^2s-\frac{y^2}{\psi}}{Z_2} ,\\
&
Z_1=(\psi+s)^2+(1-q^2)x^2,\\
&
Z_2=(\psi+q^2s)^2-(1-q^2)y^2.
    \end{split} \tag{A7}
\end{equation}

The second derivatives of the potential  $\phi(s,q)$ are

1) \begin{equation}\phi_{xx}(s,q)=\alpha_{x,x}+W_{1x}+W_{2x}, \tag{A8} \end{equation}
where

\begin{equation}\alpha_{x,x}=b \frac{\psi+s-\frac{q^2x^2}{\psi}}{Z_1}; \tag{A9} \end{equation}
\begin{equation}
     \begin{split}
&W_{1x}=\frac{b}{Z_1}\left(\psi-\frac{q^2(s^2+2x^2)}{\psi}+\frac{q^4x^2(x^2+s^2)}{\psi^3}\right)+\\
&+\frac{2bx^2}{Z_1^2}\left(-\psi+\frac{q^2x^2}{\psi}+\frac{q^2s^2}{\psi}-q^2s+\frac{q^4sx^2}{\psi^2}+\frac{q^4s^3}{\psi^2}\right);
\end{split}
\tag{A10}
\end{equation}
\begin{equation}
W_{2x}=-\frac{bq^2y^2}{Z_2}\left(\frac{1}{\psi}-\frac{q^2x^2}{\psi^3}\right)+\frac{bq^4y^2}{Z_2^2}\frac{2x^2(\psi+q^2s)}{\psi^2}; \tag{A11}
\end{equation}
\begin{equation}Z_1=(\psi+s)^2+(1-q^2)x^2, Z_2=(\psi+q^2s)^2-(1-q^2)y^2. \tag{A12} \end{equation}

2) \begin{equation}\phi_{yy}(s,q)=\alpha_{y,y}+U_{1y}+U_{2y}, \tag{A13} \end{equation}
where

\begin{equation}\alpha_{y,y}=b \frac{\psi+q^2s-\frac{y^2}{\psi}}{Z_2}; \tag{A14} \end{equation}
\begin{equation}U_{1y}=-\frac{b}{Z_1}\left( \frac{x^2+s^2}{\psi}+s-\frac{(x^2+s^2)y^2}{\psi^3}\right)+\frac{2(\psi+s)y^2b}{\psi Z_1^2}\left(\frac{x^2+s^2}{\psi}+s\right);  \tag{A15} \end{equation}
\begin{equation}U_{2y}=\frac{b}{Z_2}\left(\psi+q^2s-\frac{2y^2}{\psi}+\frac{y^4}{\psi^3}\right)-\frac{2bq^2y^2(\psi+s)}{Z_2^2}\left(1+\frac{q^2s}{\psi}-\frac{y^2}{\psi^2}\right). \tag{A16}
\end{equation}

3) \begin{equation}  \phi_{xy}(s,q)=\alpha_{x,y}+W_{1y}+W_{2y}, \tag{A17} \end{equation}
where

\begin{equation}\alpha_{x,y}=-b\frac{xy}{\psi Z_1} \tag{A18} \end{equation}
\begin{equation}W_{1y}=\frac{bxy}{Z_1}\left(\frac{1}{\psi}+\frac{q^2(x^2+s^2)}{\psi^3}\right)-\frac{2bxy(\psi+s)}{Z_1^2\psi}\left(\psi-\frac{q^2(x^2+s^2)}{\psi}\right); \tag{A19}
\end{equation}
\begin{equation} \label{secodder} W_{2y}=-\frac{bq^2xy}{Z_2\psi}\left(2-\frac{y^2}{\psi^2}\right)+2\frac{bq^4xy^3(\psi+s)}{Z_2^2\psi^2}. \tag{A20}
\end{equation}

The circular velocity in the symmetry plane of an
isothermal spheroid is defined by the formula
\begin{equation}
\begin{split}
V_c^2 (R)&=V_c^2\left(1-\frac{e}{\arcsin(e)}\frac{s}{\sqrt{R^2+e^2s^2}}\arctan\left[\frac{\sqrt{R^2+e^2s^2}}{q_3\,s}\right]\right). \end{split}
\tag{A21}
\end{equation}

\begin{center} \textbf{Model I} \end{center}

The lensing potential for model I is
\begin{equation}
\Phi_I=f_d\{\phi(0, q_{d})-\phi(a_d, q_{d}) + \phi(a_h, q_{h})\}+(1-f_d)\phi(0, q_{h}), \tag{A22}
\end{equation}
where $\phi(0, q_d)$ and $\phi(0, q_h)$ are the potentials for a singular isothermal ellipsoid with axis ratios $q_d$ and $q_h$, respectively, $\phi(a_d, q_d)$ and $\phi(a_h, q_h)$ are the potentials of an isothermal ellipsoid with cores $s=a_d$ and $s=a_h$ and axis ratios $q_d$ and $q_h$, respectively.

The first and second derivatives of the potential $\Phi_I$
can be calculated from Eqs. (A7)-(A20) using (A4).

\begin{center} \textbf{Model II} \end{center}

The lensing potential for model II is
\begin{equation}\Phi_{II}=\phi_k(a_d, q_{d})+\phi(a_h, q_{h}), \tag{A23} \end{equation} where
\begin{equation}\phi_k(s, q)=\frac{1}{2}b_k^2\ln[(\psi+s)^2+(1-q^2)x^2] \tag{A24} \end{equation}
is the potential of a Kuzmin disk, $\phi(a_h, q_h)$ ia the potential of an isothermal ellipsoid with a core $a_h$ and an axis ratio $q_h$.

$b_k$ is the normalization factor that can
be determined from the relation $m_d=\pi b_k^2\Sigma_{cr}$, where $m_d$ is the disk mass, $\Sigma_{cr}=\frac{c^2D_s}{4\pi GD_dD_{ds}}$ is the critical density, $a_d$ is the characteristic disk scale normalized
to the Einstein-Chwolson radius.

The derivatives of the Kuzmin disk potential $\phi_k(s,q)$ are

\begin{equation}\phi_k(s, q)_x=\frac{b_k^2x}{Z_{1}}L_1, \tag{A25} \end{equation}
where $L_1=\frac{\psi+q^2s}{\psi}.$

\begin{equation}\phi_k(s, q)_y=\frac{b_k^2y}{Z_{1}}L_2, \tag{A26} \end{equation}
where $L_2=\frac{\psi+s}{\psi}$.

\begin{equation} \phi_k(s, q)_{xx}=\frac{b_k^2}{Z_{1}}\left(L_1-\frac{q^4sx^2}{\psi^3}-2\frac{x^2}{Z_{1}}L_1^2\right) \tag{A27} \end{equation}

\begin{equation}\phi_k(s, q)_{yy}=\frac{b_k^2}{Z_{1}}\left(L_2-\frac{sy^2}{\psi^3}-2\frac{y^2}{Z_{1}}L_2^2\right)  \tag{A28} \end{equation}

\begin{equation}\phi_k(s, q)_{xy}=-\frac{b_k^2xy}{Z_{1}}\left(\frac{q^2s}{\psi^3}+\frac{2}{Z_{1}}L_1L_2\right)
\tag{A29}
\end{equation}

\newpage

\begin{center}\textbf{ Model III} \end{center}
The lensing potential for model III is defined by the
formula
\begin{equation}\Phi_{III}=\phi_k(a_d, q_{d})+\phi_k(a_b, q_{b})+\phi(a_h, q_{h}), \tag{A30} \end{equation}
where $\phi_k(a_d, q_d)$ è $\phi_k(a_b, q_b)$ are the potentials of a Kuzmin disk with cores $s=a_d$ è $s=a_b$ and axis ratios $q_d$ and $q_h$, respectively, $\phi(a_h, q_h)$ is the potential of an isothermal ellipsoid with a core  $a_h$ and axis ratios $q_h$.

\bigskip



\begin{references}
\reference{1.} A. D. Biggs, I. W. A. Browne, P. Helbig, et al., Mon.
Not. R. Astron. Soc. 304, 349 (1999).

\reference{2.} A. D. Biggs, I. W. A. Browne, P. N. Wilkinson, et al.,
ASP Conf. Ser. 304, 137 (2001).

\reference{3.} A. D. Biggs, O. Wucknitz, R. W. Porcas, et al., Mon.
Not. R. Astron. Soc. 338, 599 (2003).

\reference{4.} I . W. A. Browne, A. R. Patnaik, D. Walsh, and
P. N. Wilkinson, Mon. Not. R. Astron. Soc. 263, L32
(1993).
\reference{5.} A. S. Cohen, J. N. Hewitt, C. B. Moore, and
D. B. Haarsma, Astrophys. J. 545, 578 (2000).

\reference{6.} W. Dehnen and J. Binney, Mon. Not. R. Astron. Soc.
294, 429 (1998).

\reference{7.} C. Feron, J. Hjorth, J. P. McKean, and J. Samsing,
Astrophys. J. 696, 1319 (2009).

\reference{8.} R. de Grijs and P. C. van der Kruit, Astron. Astrophys.
Suppl. Ser. 117, 19 (1996).

\reference{9.} H. J. Grimm, M. Gilfanov, and R. Sunyaev, Astron.
Astrophys. 391, 923 (2002).

\reference{10.} J. N. Hewitt, E. L. Turner, D. P. Schneider, et al.,
Nature 333, 537 (1988).

11. J. Holmberg and C. Flynn,Mon. Not. R. Astron. Soc.
352, 440 (2004).

12. A. Kassiola and I. Kovner, Astrophys. J. 417, 450
(1993).

13. C. R. Keeton and C. S. Kochanek, Astrophys. J. 495,
157 (1998).

14. D. D. Kelson, G. D. Illingworth, P. G. van Dokkum,
and M. Franx, Astrophys. J. 531, 159 (2000).

15. I . R. King, An Introduction to Classical Stellar Dynamics
(Univ.of California, Berkeley, 1994; Editorial
URSS,Moscow, 2002).

16. R. Kormann, P. Schneider, and M. Bartelmann, Astron.
Astrophys. 284, 285 (1994).

17. G. G. Kuzmin, Astron. Zh. 33, 27 (1956).

18. R. Mittal, R. Porcas, O. Wucknitz, et al., Astron.
Astrophys. 447, 515 (2006).

19. S. Nair, D. Narasimha, and A. P. Rao, Astrophys. J.
407, 46 (1993).

20. R. Narayan andM. Bartelmann, astro-ph/9606001v2
(1997).

21. A. Patnaik, I. Browne, L. King, et al., Mon. Not.
R. Astron. Soc. 261, 435 (1993).

22. A. R. Patnaik, R.W. Porcas, andW. A. Browne,Mon.
Not. R. Astron. Soc. 274, L5 (1995).

23. H. J. Reid and A. Brunthaler, Astrophys. J. 616, 872
(2004).

24. P. D. Sackett, Astrophys. J. 483, 103 (1997).

25. M. Schmidt, Bull. Astron. Inst. Netherlands 13, 15
(1956).

26. P. Schneider, J. Ehlers, and E. E. Falco, Gravitational
Lenses, 2nd ed. (Springer, Berlin, Heidelberg,
New York, Barcelona, Hong Kong, London, Milan,
Paris, Singapore, Tokio, 1999).

27. Y. Sofue and V. Rubin, Ann. Rev. Astron. Astrophys.
39, 137 (2001).

28. O.Wucknitz, A. D. Biggs, and I . W. A.Browne,Mon.
Not. R. Astron. Soc. 349, 14 (2004).

29. T.York, N. Jackson, I .W. A. Browne, et al.,Mon.Not.
R. Astron. Soc. 357, 124 (2005).
\end{references}
\end{document}